# Field effect-induced tunability in planar hyperbolic metamaterials

Georgia T. Papadakis[1,†] and Harry A. Atwater[1]

[1] Thomas J. Watson Laboratories of Applied Physics, California Institute of Technology, California 91125, USA

[†]Corresponding author e-mail: gpapadak@caltech.edu

**Abstract**

We demonstrate that use of the field effect to tune the effective optical parameters of a layered hyperbolic metamaterial leads to topological transitions in its dispersion characteristics in the optical regime. Field effect gating electrically modulates the permittivity in transparent conductive oxides via changes in the carrier density. These permittivity changes lead to active extreme modulation of ~200% of the effective electromagnetic parameters along with active control of the anisotropic dispersion surface of hyperbolic metamaterials and enable the opening and closing of photonic band gaps.

## I. INTRODUCTION

Over the past decade, extensive study of optical metamaterials, artificial materials composed of sub-wavelength nanostructures, has revealed light-matter interactions unseen in the natural world. Their exotic properties arise from engineering their effective optical parameters, namely, their effective electric permittivity and magnetic permeability, to be radically different from those of natural materials. Recently, epsilon near zero (ENZ)[1,2] and mu near zero[3] metamaterials have been reported as slow-light media, negative index materials that enable super-resolution imaging[4,5] have been demonstrated and high-index metamaterials for sub-diffraction imaging[6] have been investigated. Hyperbolic metamaterials are a special class of metamaterials with uniaxial anisotropy and effective parameters of opposite signs along different coordinate directions[7,8]. Such systems are being intensively studied because they display diverging density of states, enabling extreme Purcell factors to be achieved[9], and support negative refraction of power[7]. They are also being investigated for emissivity control engineering for solar applications[10,11]. The range of fundamental phenomena and applications achievable by metamaterials can be significantly expanded by actively tuning their effective electromagnetic parameters to enable dynamic control over their optical response. Such active control over metamaterials intrinsic properties can pave the way towards novel active optical components like holographic displays, tunable polarizers, sensors and switches, slow-light media and optical memories.

Tuning the effective electromagnetic parameters of metamaterials also leads to active control over their dispersion characteristics. The past decade has revealed a plethora of new phenomena both in electronic materials and condensed matter physics and their photonic counterparts arising from engineering their electronic band structure and photonic dispersion surface respectively. Classical phenomena such as Lifshitz transitions[12] and Van Hove singularities that lead to extreme values of magnetoresistance arise from inducing transitions in the Fermi surface of electronic systems. Quantum-mechanical effects such as those arising from Dirac-like dispersion surfaces, leading to topological Dirac phases are the result of band structure engineering, leading to three-dimensional semimetals[13] and topological insulators[14]. Photonic analogues of topological insulators, in other words photonic materials that support non-trivial topologically protected states against back-scattering, have been realized with helical waveguides in a honeycomb lattice[15], by engineering the dispersion characteristics of photonic crystals. Theoretical studies suggest other configurations for realization of such protected states like chiral metamaterials with equal values of electric permittivity and magnetic permeability[16], or with chiral hyperbolic metamaterials[17] or with index-near-zero (which can be generalized to effective electric permittivity or magnetic permeability near-zero) metamaterials[18, 19]. The latter idea arises from the simple consideration that Dirac-like cones of the dispersion can be realized in a linear dispersion regime where the effective parameters of the metamaterial approach zero, allowing for topological transitions to occur in metamaterials. Hyperbolic metamaterials support ENZ and epsilon near pole (ENP) spectral regions[20, 21] that induce band splitting and support such topological transitions[22]. Thus, active tuning of the effective parameters can tune the spectral regions of topological transitions along with the possibility to dynamically study topological transitions in photonic systems.

Potential tuning mechanisms include modifying the complex dielectric function of component materials via phase transitions[23], mechanical deformations[24] and electronic mechanisms[25, 26] for tuning such as charge carrier injection, accumulation and depletion[27]. Among these, field effect modulation is particularly attractive because of its robustness and low power dissipation in steady state. It has recently been investigated for optical modulators[28-32] by using the spectral tunability of the electronic properties in transparent conductive oxides (TCOs) for modulating the modal effective index in waveguide configurations.

Here we investigate the effects of carrier density changes in TCO accumulation layers on the effective electromagnetic optical parameters of hyperbolic metamaterials. We investigate a frequency and dispersion-tunable hyperbolic metamaterial (HMM) with field effect gating to electrically modulate the permittivity in conductive oxides layers. We find that such modulation can induce transitions in the metamaterial dispersion surface from elliptical to hyperbolic, which is identified as a topological transition of the dispersion surface[22], leading to singularities in the density of optical states. We calculate the effective electric permittivity and magnetic permeability tensors of the HMM and we show that active permittivity modulation can be as high as 200% for experimentally realizable geometries and TCOs. This also results in tunability of the effective birefringence and dichroism of the metamaterial and in tunability of its optical band gaps. We perform a sensitivity analysis of the electronic properties of the TCO, to provide insight for practical realization of tunable HMMs with various conductive oxides and degenerately doped semiconductors. The design simplicity and robustness of field effect modulation, similar to structures now widely used in CMOS electronics, opens the door to applications in future optical technologies.

## II. TUNABLE PLANAR HYPERBOLIC METAMATERIAL: THE CONCEPT

A hyperbolic metamaterial can be realized in a stack of metal-dielectric multilayers with sub-wavelength thicknesses. It is characterized through a tensorial electric permittivity $\vec{\varepsilon} = diag(\varepsilon_{xx}, \varepsilon_{yy}, \varepsilon_{zz}) = diag(\varepsilon_o, \varepsilon_o, \varepsilon_e)$ and magnetic permeability $\vec{\mu} = diag(\mu_{xx}, \mu_{yy}, \mu_{zz}) = diag(\mu_o, \mu_o, \mu_e)$, where $\varepsilon_o$ and $\mu_o$ refer to the ordinary parameters in the in-plane direction, while $\varepsilon_e$ and $\mu_e$ correspond to the direction of the optical axis. The strong anisotropy along the optical axis causes the TM polarized bulk modes, containing an electric field component in this direction, to experience strong birefringence expressed via the dispersion equation[20]:

$$\frac{k_x^2 + k_y^2}{\varepsilon_e \mu_o} + \frac{k_z^2}{\varepsilon_o \mu_o} = k_o^2 \qquad (1)$$

Recent studies[20, 21] have shown that planar HMMs can be engineered to exhibit hyperbolic dispersion of type I ( $\mathrm{Re}(\varepsilon_e \mu_o) < 0$ and $\mathrm{Re}(\varepsilon_o \mu_o) > 0$ ) and type II ( $\mathrm{Re}(\varepsilon_e \mu_o) > 0$ and $\mathrm{Re}(\varepsilon_o \mu_o) < 0$ ), in addition to regions where propagation is forbidden due to simultaneously negative values of the denominators in Eq. 1. By introducing a TCO within the unit cell of a metal/dielectric metamaterial, as shown in Fig. 1, and actively inducing changes in carrier density through application of a DC bias across the metal/TCO layers, a carrier accumulation layer is formed at the dielectric/TCO interface, shown with light blue color in Fig. 1.

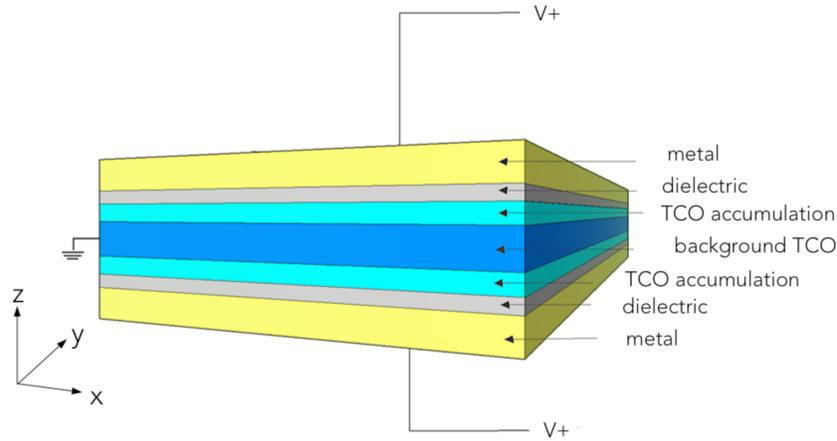

**Fig.1**: Schematic of a unit cell of a tunable planar metamaterial.

This changes Maxwell's boundary conditions on those interfaces, which affects the effective metamaterial parameters, which are obtained through a parameter retrieval method[20]. The optical parameter affected the most through this modulation is the extraordinary permittivity $\varepsilon_e$ as we show in Part B. This influences the metamaterial band structure, birefringence and dichroism. Prior to investigating a specific unit cell we discuss possible dielectrics and TCOs that can be realistically implemented in a tunable HMM unit cell like the one shown in Fig. 1. Specifically, for inducing strong tunablility, we investigate high-strength dielectrics for sustaining large breakdown fields and TCOs with high background carrier concentrations that translate to plasma frequencies in the near-IR regime.

### A. ELECTRONIC PROPERTIES: HIGH-STRENGTH DIELECTRICS AND TCOs

Ignoring detailed band bending effects, to first order, charge carrier accumulation can be schematically modeled via a uniform layer with increased carrier concentration relative to the background TCO, with thickness given by the Debye length $d$. A simple electrostatic calculation on a metal/dielectric/TCO interface, as depicted in Fig. 1, dictates that:

$$\frac{V}{t} = \frac{|e|Nd}{k_{diel}\varepsilon_o} \leq E_{br} \tag{2}$$

where $V$ is the applied bias between the metal and TCO, $t$ is the thickness of the dielectric, $k_{diel}$ is the DC dielectric constant of the dielectric, $\varepsilon_o$ is the vacuum permittivity and $E_{br}$ is the breakdown field of the dielectric material. $N$ is the carrier density in the accumulation layer. To achieve tunability of the effective parameters within the visible regime, the TCO incorporated into the metamaterial design must have a plasma frequency within the IR-visible spectrum. This corresponds to high background carrier concentration in the order of $10^{19}-10^{20}/cm^3$ and it dictates the use of dielectric materials with high electrical breakdown fields.

The TCO we investigate in this work is indium tin oxide (ITO), which has been studied extensively in the literature[28, 29, 33-36]. We model ITO with a Drude model, which provides an adequate and accurate description of its permittivity in the visible-IR regime[28, 29, 34, 36, 37]. ITO can be heavily doped using rf sputter deposition, yielding a carrier concentration in the range $10^{19}-10^{21}/cm^3$ with a plasma frequency in the infrared range. Following previous experimental results [28, 29, 34, 38], we investigate here ITO with background carrier concentration of $5\times10^{20}/cm^3$. Active tunability of the carrier concentration of ITO in metal-oxide-semiconductor (MOS) structures by as much as two orders of magnitude has been reported previously[28,29,30, 35, 36]. The results presented here are also relevant for other TCOs like aluminum doped zinc oxide (AZO)[33], gallium doped zinc oxide (GZO)[32, 33] or degenerately doped semiconductors [31], as illustrated by a sensitivity analysis on the material electronic properties of the active layer discussed in Part II, C.

Table I illustrates the maximum achievable carrier concentration in the accumulation layer of ITO, for various dielectric materials whose breakdown fields have been reported: $SiO_2$, $Al_2O_3$, $HfO_2$ and $HfSiO_4$. Different Debye lengths $d$ for the ITO are also considered, consistent with the range of parameters observed in previous experimental results [28-30, 35].

| dielectric | $E_b$ and and $k_{diel}$ | N (in cm$^{-3}$), d=0.5nm | N (in cm$^{-3}$), d=1nm | N (in cm$^{-3}$), d=2.5nm | N (in cm$^{-3}$), d=5nm |
|---|---|---|---|---|---|
| $SiO_2$ | $E_b$=[30-40] MV/cm [39]<br>$k_{diel}$=3.9 [39-41] | [1.29-1.7]x10$^{22}$ | [6.47-8.63]x10$^{21}$ | [2.59-3.45]x10$^{21}$ | [1.29-1.73]x10$^{21}$ |
| $Al_2O_3$ | $E_b$=[6-8] MV/cm [40, 42, 43]<br>$k_{diel}$=9 [40, 41] | [5.97-7.96]x10$^{21}$ | [2.99-3.98]x10$^{21}$ | [1.19-1.59]x10$^{21}$ | [5.97-7.96]x10$^{20}$ |
| $Al_2O_3$ | $E_b$=[6-8] MV/cm [40, 42, 43]<br>$k_{diel}$=10.3 [43] | [6.83-9.12]x10$^{21}$ | [3.42-4.56]x10$^{21}$ | [1.37-1.82]x10$^{21}$ | [6.84-9.12]x10$^{21}$ |
| $HfO_2$ | $E_b$=[40-60] MV/cm [39]<br>$k_{diel}$=17 [39] | [7.52-11.28]x10$^{22}$ | [3.76-5.64]x10$^{22}$ | [1.5-2.26]x10$^{22}$ | [7.52-11.28]x10$^{21}$ |
| HfO$_2$ | E$_b$=[40-60] MV/cm [39] k$_{diel}$=25 [41] | [11.1-16.6]x10$^{22}$ | [5.53-8.23]x10$^{22}$ | [2.21-3.32]x10$^{22}$ | [1.11-1.66]x10$^{22}$ |
| HfO$_2$ | E$_b$=5.6 MV/cm [43] k$_{diel}$=18.7 [43] | 1.55x10$^{22}$ | 7.74x10$^{21}$ | 3.09x10$^{21}$ | 1.55x10$^{21}$ |
| HfSiO$_4$ | E$_b$=10 MV/cm [41,44] k$_{diel}$=11 [41,44] | 1.22x10$^{22}$ | 6.08x10$^{21}$ | 2.43x10$^{21}$ | 1.22x10$^{21}$ |
| HfSiO$_4$ (with SiO$_2$) | E$_b$=10 MV/cm [41,44] k$_{diel}$=[4.8-5.4] [45] | [5.31-5.97]x10$^{21}$ | [2.66-2.99]x10$^{21}$ | [1.06-1.19]x10$^{21}$ | [5.31-5.97]x10$^{20}$ |
| HfSiO$_4$ (with HfO$_2$) | E$_b$=10 MV/cm [41,44] k$_{diel}$=[12.5-15.1] [45] | [1.38-1.67]x10$^{22}$ | [6.91-8.35]x10$^{21}$ | [2.77-3.34]x10$^{21}$ | [1.38-1.67]x10$^{20}$ |

**Table.1**: Maximum achievable carrier concentration in ITO accumulation layers for reported values of breakdown field and DC dielectric constants of high-strength dielectrics

HfO$_2$ has the highest breakdown field and DC dielectric constant, which suggests that it may be the most suitable dielectric for sustaining high carrier concentrations at the accumulation layer of ITO, and in turn provides large tunability of the effective parameters of a planar HMM.

### B. A 'TOY' MODEL: RESULTS

We consider a unit cell of a HMM consisting of two 20 nm Ag layers separated by a 15 nm layer of ITO. The two materials are isolated from each other by 10 nm HfO$_2$ dielectric layers, as shown in Fig. 1. The thickness of HfO$_2$ has been chosen to correspond to thicknesses routinely achievable in atomic layer deposition[42,43]. The refractive index of Ag has been measured by ellipsometry (See Supplemental Material), while the Sellmeier equation with three poles was used for the refractive index of HfO$_2$ [46]. Under applied bias between the Ag and the ITO in the geometry of Fig. 1, we consider a 2.5nm accumulation layer to be formed in the HfO$_2$-ITO interface[28-30]. In Part II, C, we perform a sensitivity analysis over the accumulation layer thickness in order to comprehensively assess the tunability range for this design for various TCOs and degenerate semiconductors. We model both the background ITO and the accumulation layer using the Drude model with different carrier concentration, based on previous experimental results[28,29,34,35] (See Supplemental Material).

The drastic change in the carrier concentration across the ITO background and accumulation layer yields tunable optical parameters. We use a parameter retrieval method[20] to calculate the effective optical parameters $\varepsilon_o$, $\mu_o$ and $\varepsilon_e$ and $\mu_e$ of this motif for increasing carrier concentration in the accumulation layer of ITO. The retrieved parameters are local and thus independent of the number of unit cells (See Supplemental Material).

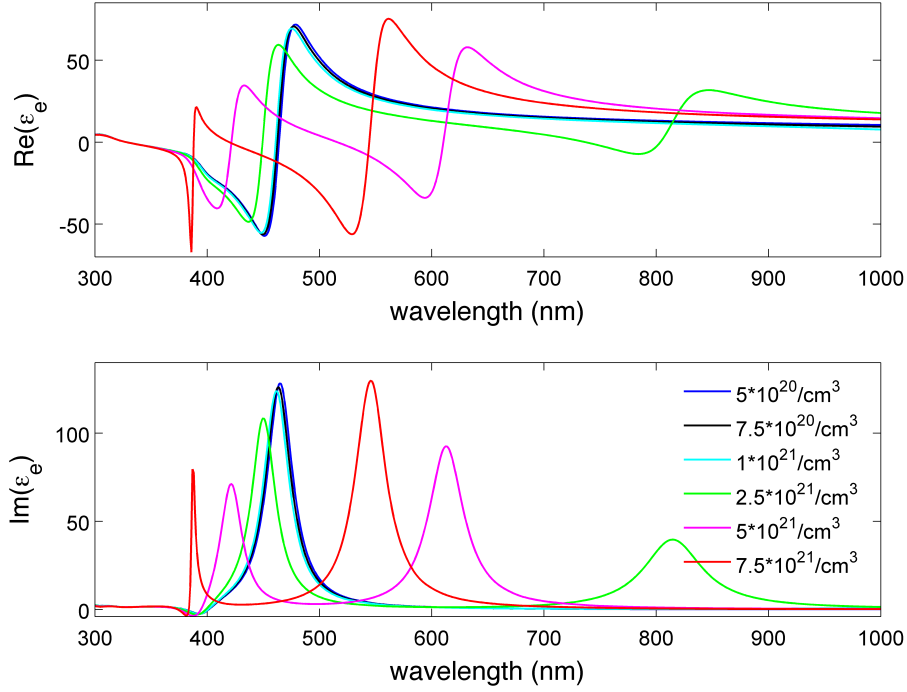

**Fig. 2**: Tunable effective extraordinary permittivity for various carrier concentrations in the accumulation layer

As the carrier concentration in the accumulation layer of ITO increases under applied bias from a background value of $5\times10^{20}/cm^3$ up to $7.5\times10^{21}/cm^3$, corresponding to experimentally reported carrier concentration changes[28], the Lorentzian-shaped resonance in extraordinary permittivity $\varepsilon_e$ blue-shifts as seen in Fig. 2. This resonance arises form the coupling of the plasmonic modes supported on the metal/dielectric interfaces to the bulk high-k modes of the HMM[21]. The epsilon-near-pole (ENP) wavelength of this dominant resonance blue-shifts by more than 60nm while remaining within the visible regime under applied bias. More importantly, as depicted in Fig. 2, the application of a DC bias across the metamaterial layers also leads to the appearance of a secondary resonance, originating from the large increase in the charge density in the ITO accumulation layer carrier density.

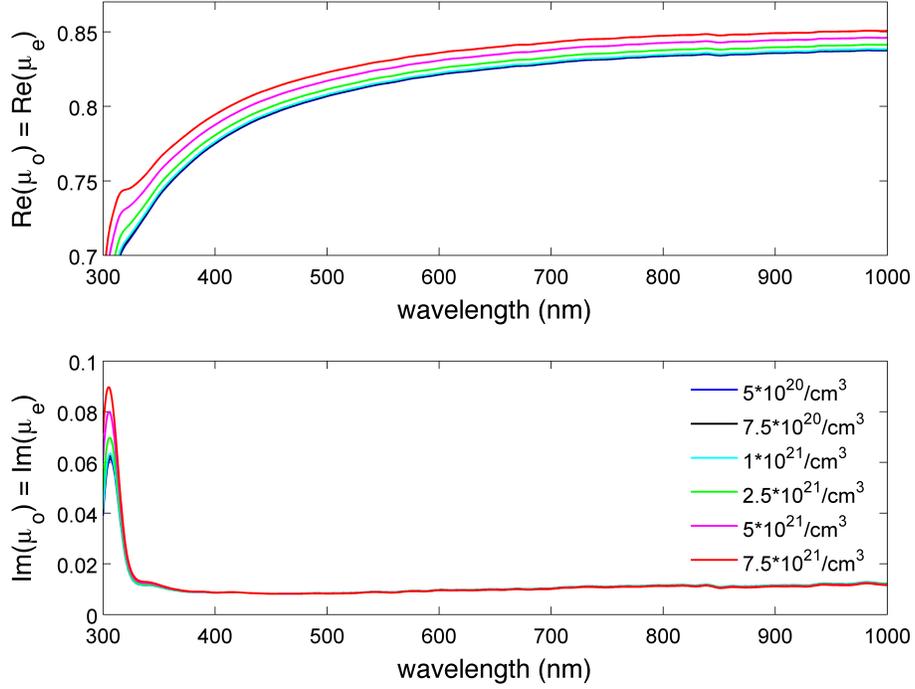

**Fig. 3**: Tunable effective extraordinary permeability for various carrier concentrations in the accumulation layer

The HMM unit cell exhibits a strong and broadband diamagnetic response. One of the most diamagnetic natural material is bismuth with a permeability of 0.999834 while this design enables permeability values in the order of 0.85. This is a consequence of Lenz's law: the tangential component of the magnetic field induces a surface current at the interfaces of the multilayer, creating a magnetic response opposing the applied magnetic field. The magnetic permeability, which is isotropic, i.e $\mu_o = \mu_e$, is also noticeably tunable, as shown in Fig. 3. Since this HMM is magnetically isotropic, the TE polarized bulk modes do not experience any anisotropy since they are only affected by the ordinary permittivity $\varepsilon_o$, which is known to be metallic-like in planar HMMs[20]. The retrieved $\varepsilon_o$ is not significantly tunable with variation of carrier concentration (See Supplemental Material) since the in-plane response of this HMM is dominated by Ag layers rather than tunable ITO accumulation layers.

By contrast, the TM polarized bulk modes experience a strong anisotropy. According to Eq. 1, their effective birefringence and dichroism are defined as $\mathrm{Re}(\sqrt{\varepsilon_o \mu_o}) - \mathrm{Re}(\sqrt{\varepsilon_e \mu_o})$ and $\mathrm{Im}(\sqrt{\varepsilon_e \mu_o}) - \mathrm{Im}(\sqrt{\varepsilon_o \mu_o})$ respectively and are presented in Fig. 4 for increasing carrier concentration in the accumulation layer of ITO. As expected, HMMs exhibit extreme anisotropies, which is manifest in the large birefringence values shown in Fig. 4; these by far exceed the birefringence of the most anisotropic natural materials like liquid crystals and other uniaxial inorganic crystals (c.f., table in Fig. 4). Notable here is the broadband tunability of both birefringence and dichroism across the whole visible spectrum.

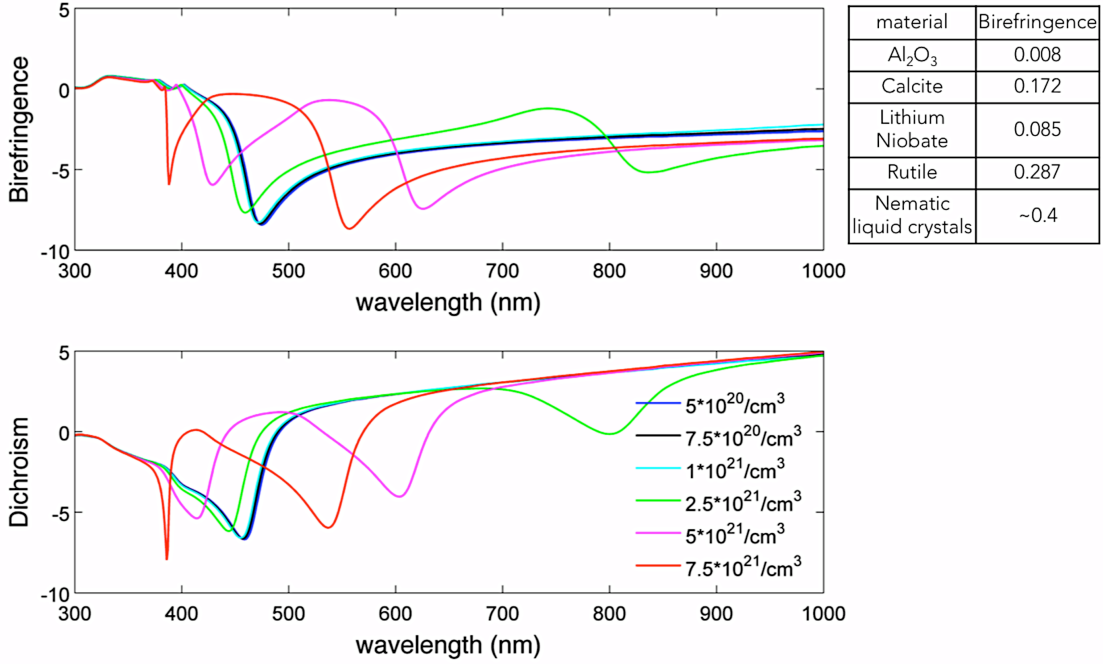

**Fig. 4**: Tunable effective birefringence and dichroism

The tunability of the effective parameters presented in Figs. 2 and 3 has an effect on the dispersion surface for the TM waves, given by Eq. (1). This is illustrated in Fig. 5 where we depict the isofrequency contours in the lossless limit for the metamaterial under consideration. Both $k_x$ and $k_z$ are normalized to the free space wavenumber $k_0$. Application of DC bias across the HMM unit cell is seen to yield drastic changes in both the shape and type of the dispersion surface. Additionally, the surface area enclosed by the isofrequency contours, which is proportional to the total number of available optical states[47] changes significantly with changes in ITO accumulation layer carrier concentration, yielding a route to active control over the metamaterial optical density of states.

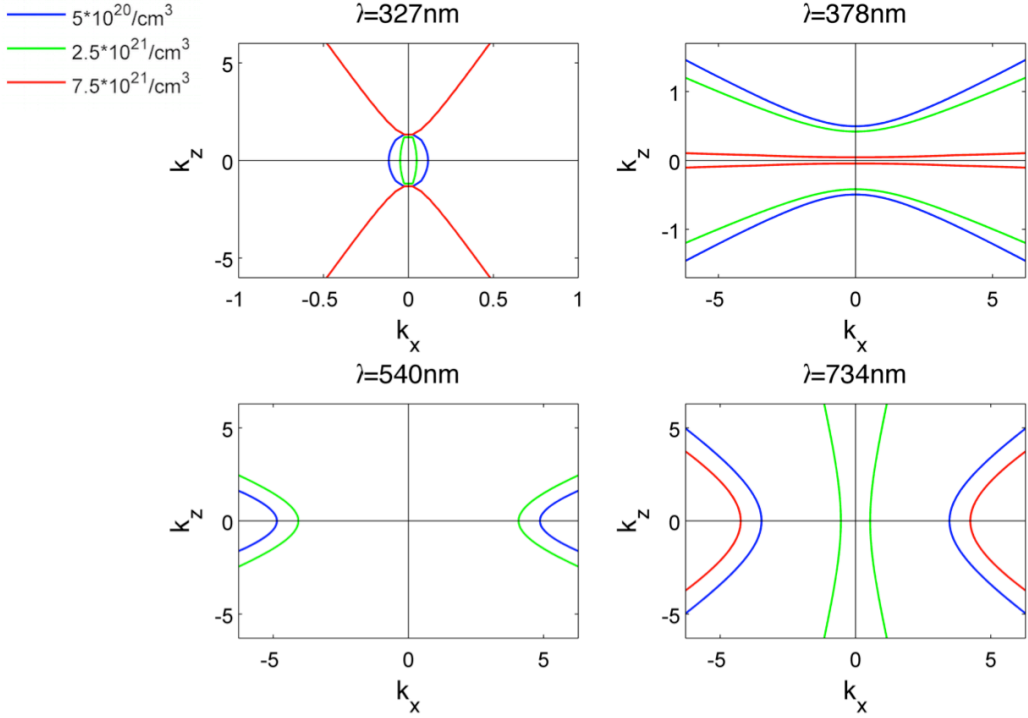

**Fig. 5**: Isofrequency contours for three different carrier concentrations in the accumulation layer in the lossless limit

For example, for a wavelength of 327nm as shown in Fig. 5: the static case of the metamaterial at 0V applied bias (blue color) exhibits an elliptical dispersion, which remains the case for small changes in the carrier density of the ITO accumulation layer (green color). However, at higher concentrations (red color), the dispersion surface undergoes a transition to a type I hyperbolic dispersion with corresponding negative values of the extraordinary permittivity $\varepsilon_e$. As the wavelength increases, the HMM supports type I hyperbolic dispersion for all the considered carrier concentrations in the ITO accumulation layer. This is a consequence of the effective built-in dipole moment in the direction across the optical axis for oblique incidence and TM polarization illumination. At the wavelength of 540nm, the absence of an isofrequency curve at the carrier concentration of $7.5 \times 10^{21}/cm^3$ is a signature of both parameters $\varepsilon_o \mu_o$ and $\varepsilon_e \mu_o$ being negative, which leads to a forbidden band where real valued wavenumbers are not allowed. This corresponds to a spectral region of an effective omnidirectional band gap[20]. In the longer wavelength regime, the isofrequency curves exhibit type II hyperbolic dispersion.

In Fig. 6, we present the normal wavenumber supported by the metamaterial for carrier concentrations $5 \times 10^{20}/cm^3$, $2.5 \times 10^{21}/cm^3$ and $7.5 \times 10^{21}/cm^3$, extracted from Eq. (1). As recently discussed in Ref. [48], the "valleys" and "ridges" of the imaginary part of the wave vector are a good criterion for identifying regions of enhanced and suppressed density of optical states respectively. Fig. 6 depicts those "valleys" and "ridges". However, we note that HMMs support high-k modes, so a large imaginary part of the wavenumber may not necessarily imply a reflective region with vanishing density of optical states, if not accompanied by a vanishing real part of the wavenumber. Thus, we include in the diagrams of Fig. 6 the real part of the wavenumber.

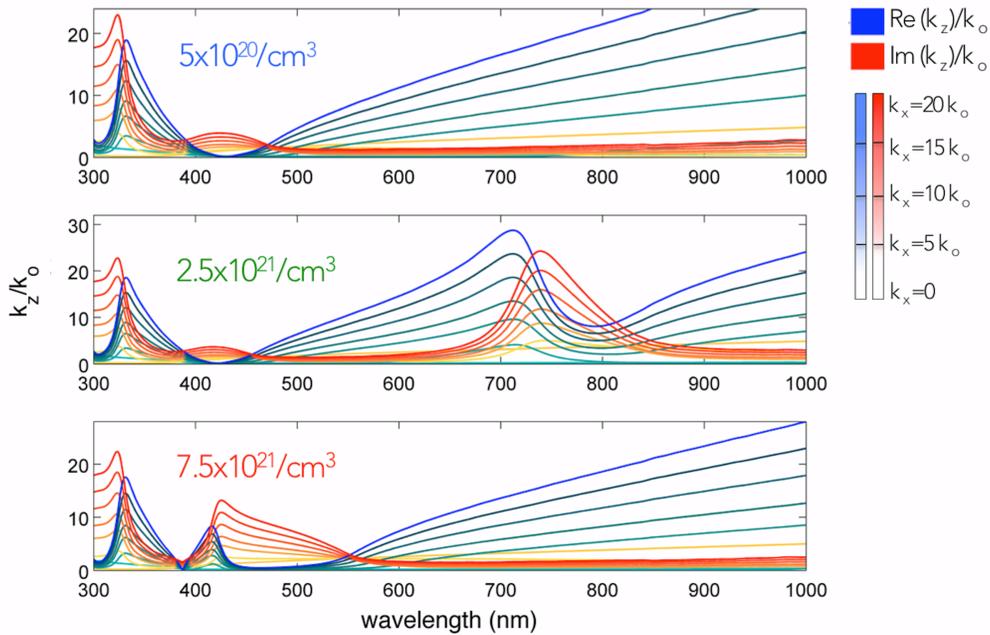

**Fig. 6**: Real (blue) and imaginary (red) part of effective normal k-vector in the hyperbolic metamaterial, for three different carrier concentrations in the accumulation layer

At 0V (carrier concentration $5 \times 10^{20} / cm^3$) the metamaterial supports three distinct spectral regions, as shown in Fig. 6, top. In the short wavelength regime between 300nm and 400nm, the real part of the wavenumber exceeds the imaginary part, leading to type I hyperbolic dispersion. This region adjoins one in which no photonic states are supported, corresponding to an omnidirectional reflective gap and suppressed density of states, since the real part of the wavenumber vanishes while the imaginary part is drastically increased. The long wavelength regime exhibits a type II hyperbolic dispersion region, as expected for planar HMMs. However, the situation is substantially altered when the carrier concentration increases. As depicted in the bottom two plots of Fig. 6, the topology of the curves representing both real and imaginary parts of the wavenumber are strongly influenced by the carrier concentration change in the ITO accumulation layer. A new effective band gap is introduced in the longer wavelength limit between 740nm and 810nm for carrier concentration of $2.5 \times 10^{21} / cm^3$. This is a direct consequence of the introduction of a secondary resonance in the extraordinary permittivity $Re(\varepsilon_e)$-(c.f., Fig. 2). This band gap is then drastically blue-shifted for higher ($7.5 \times 10^{21} / cm^3$ case) accumulation carrier concentration. This is accompanied by strong suppression of the central hyperbolic region. The figure of merit (FOM), defined as the real over the imaginary part of the normal wavenumber reaches values as high as 15 for the hyperbolic regions (as c.f., Fig. 14 of the Supplemental Material). This value by far exceeds the typically observed FOM which is reported to range between 1-3 for other metamaterial types[49], highlighting the potential of a hyperbolic dispersion for useful, physically realizable metamaterials without excessive losses[50]. By contrast, the FOM vanishes within the band gaps (c.f., Fig. 14 in supplemental material). To illustrate the effects of tunable optical parameters on the effective band structure, in the lossless limit, we show the change in the topology of the frequency-dependent three-dimensional dispersion surface for the two extreme cases of ITO accumulation carrier concentration.

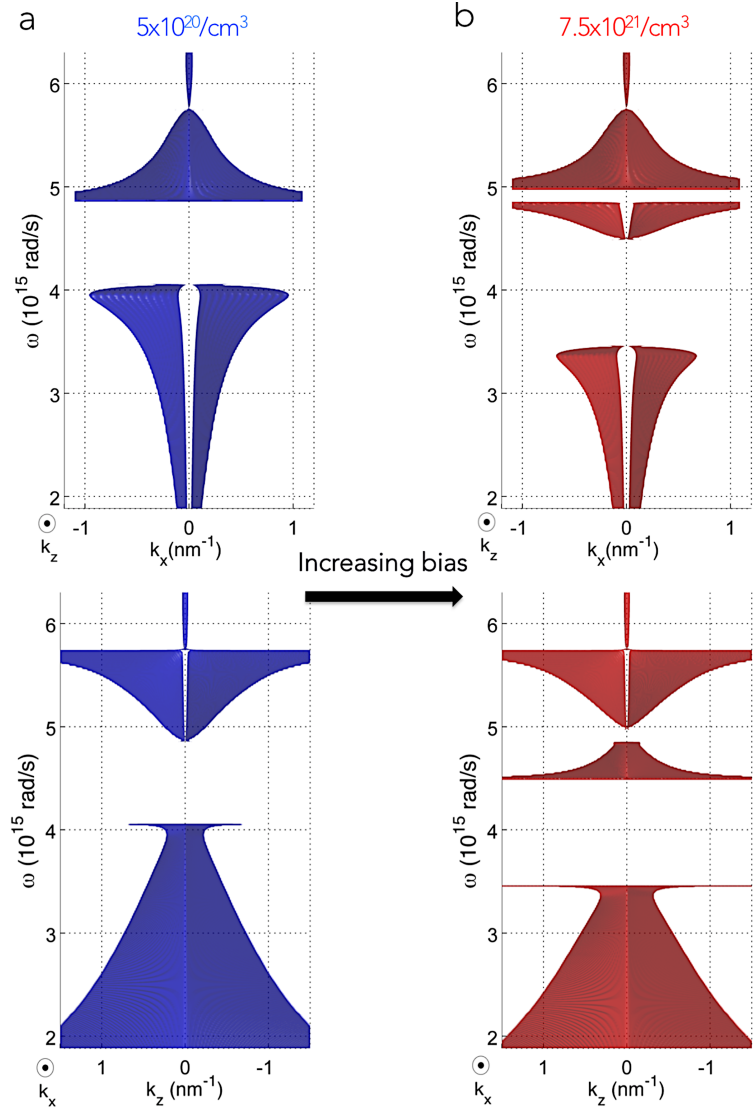

**Fig.7:** Transition of the three-dimensional dispersion surface along $k_z$ axis (top) and along $k_x$ axis (bottom) for carrier concentration in the ITO accumulation layer a) $5 \times 10^{20} / cm^3$ corresponding to 0V applied bias and b) $7.5 \times 10^{21} / cm^3$

Fig. 7 illustrates the effect of ITO electric permittivity modulation on the metamaterial dispersion surface when applying a DC bias across the Ag-HfO$_2$-ITO layers. Within the optical regime, we see an effective omnidirectional band gap whose band edges can be tuned by varying the applied bias.

Additionally, we notice the appearance of a new effective band gap and a new hyperbolic region, for larger accumulation layer carrier concentration changes. This reveals that the field effect provides sufficient change of the conductive oxide permittivity to allow for spectral shifting of the hyperbolic regions and band gaps of HMMs and even to allow for active "opening" and "closing" of band gaps.

**C. Sensitivity analysis over TCO electronic parameters**

As shown in Eq. (2), the parameter that principally defines the tunability range of the effective optical parameters of our metamaterial is the product $N \cdot d$, where $N$ is the maximum achievable carrier concentration in the accumulation layer of the TCO before electrical breakdown occurs, and $d$ is the Debye length or thickness of the accumulation layer. We perform a sensitivity analysis of the retrieved extraordinary permittivity $\varepsilon_e$, separately as a function of $N$ and $d$, since $\varepsilon_e$ is the optical parameter most drastically affected by the field effect. Both for $N$ and $d$, we consider values within the range of previous experimental reports[28-30, 34, 35, 36]. We suggest that this approach can provide useful insight here, and also for other semiconductors and TCOs as active constituent materials in field-effect-tunable metamaterial realizations. In Fig. 8 we illustrate the tunability of the ENP region of the primary Lorentzian resonance.

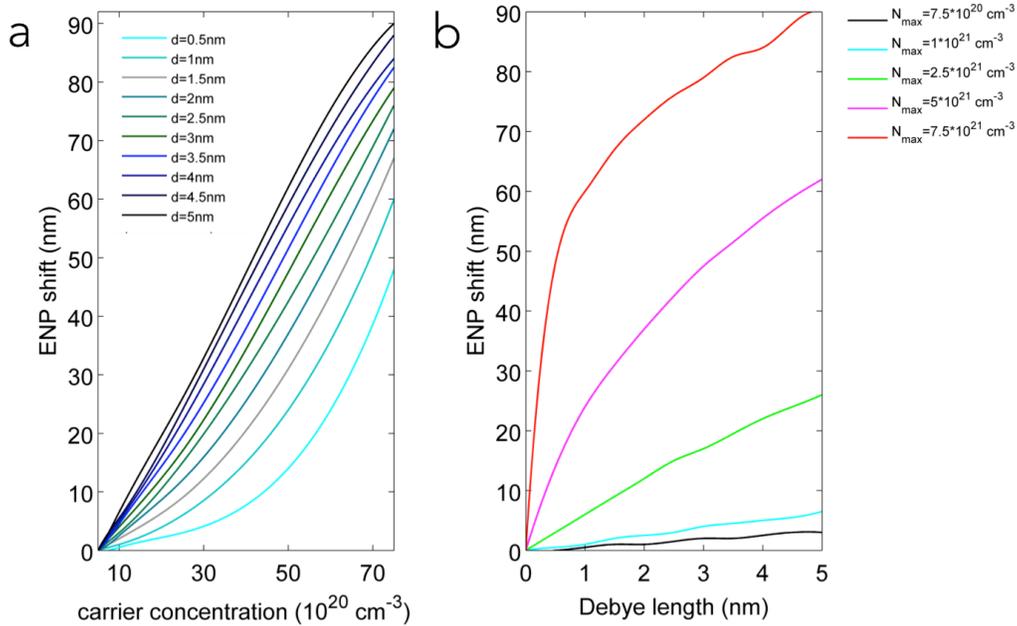

**Fig.8:** Effects of the electronic properties of ITO on the ENP wavelength of $\varepsilon_e$. a) carrier concentration change effect for varying Debye lengths b) Debye length effect for varying maximum achievable carrier concentration in the accumulation layer before breakdown occurs

The maximum achievable change in the carrier concentration of the accumulation layer before breakdown occurs is a critical factor for the ENP shift, as shown in Fig. 8a. Specifically, for experimentally realizable TCOs[28, 29, 33, 34], that have been shown to undergo large changes in carrier concentration, the wavelength shift is large enough to be experimentally detectable. Tunability of the ENP wavelength for metamaterials that undergo an elliptical to hyperbolic transition, should give rise to readily observable changes in photoluminescence[51] or cathodoluminescence[52, 53] intensities.

As indicated from Fig. 2, it is the appearance of secondary resonances in $\varepsilon_e$, upon application of DC bias across the metamaterial layers that has the strongest effect on tuning the birefringence, dichroism and effective band structure of the metamaterial. To investigate this, we calculate below the effect of the Debye thickness and maximum achievable carrier concentration change on the relative change of the real part of $\varepsilon_e$, $|\Delta \varepsilon_e|_{rel} = |\varepsilon_{e\_0V} - \varepsilon_e| / \varepsilon_{e\_0V}$ under applied bias.

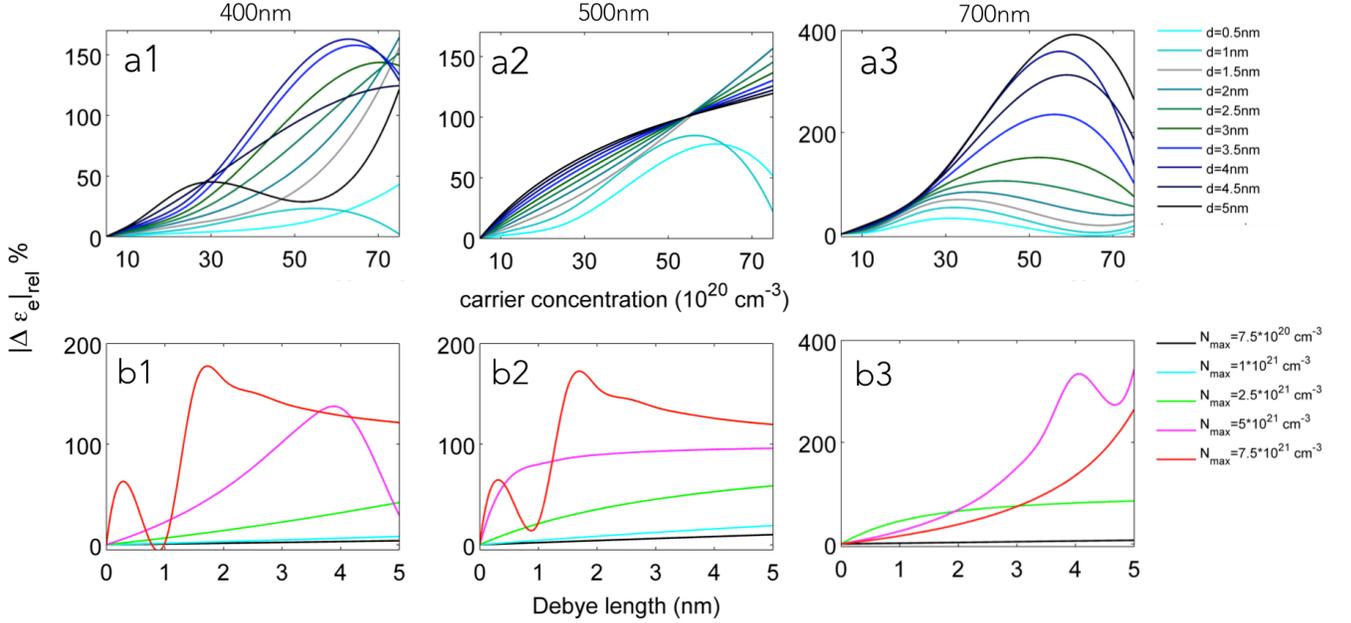

**Fig.9:** Variation in the electronic properties of ITO and their effect on relative change of $\varepsilon_e$. Top: carrier concentration change effect for varying Debye lengths for wavelengths of a1) 400nm, a2) 500nm, a3) 700nm and bottom: Debye length effect for varying maximum achievable carrier concentration in the accumulation layer before breakdown occurs for b1) 400nm, b2) 500nm, b3) 700nm

As shown in Figs. 9, a1, 2 and 3, even for Debye lengths in the order of 0.5nm, the relative change in the value of $\varepsilon_e$ is in the order of 20%-100% for the wavelengths accessed here. This large change arises from the appearance of secondary resonances in $\varepsilon_e$. Figs. 9, b1, 2, and 3 suggest that achieving very high carrier densities in the accumulation layer of the TCO facilitates the strong, experimentally observable tunability effects. For example, as seen from the red and magenta curves in Figs. 9 b1, 2, 3, the relative change $|\Delta\varepsilon_e|_{rel}$ can be as high as 150%-350% for a 100x modulation of the TCO accumulation carrier density.

### III. CONCLUSION

In conclusion, we have outlined a method for electronically tuning the topology of the dispersion surface of hyperbolic metamaterials by field-effect gating and electrical modulation of the permittivity in transparent conductive oxide layers. We observe the "opening" and "closing" of omnidirectional band gaps controlled by applied bias, which corresponds to a tunable figure of merit, with values as high as 15 in the hyperbolic regime while vanishing at the band gaps. The field-effect induced changes in the permittivity of the TCO accumulation layers give rise to spectral tunability of the effective parameters of hyperbolic metamaterials. We observe blue-shifts of the effective permittivity along the optical axis by more than 60nm in the visible regime and additional resonances are introduced due to the large changes in carrier density across metal-dielectric-TCO triplet layers. This gives rise to broadband tunability of the effective birefringence and dichroism, with potential for novel photonic devices like tunable metal-dielectric waveguides, optical sensors, filters and polarizers. Such active control over the complex parameters of metamaterials is also essential for slow light media and holographic display realizations. A sensitivity analysis of the extraordinary permittivity to changes in the TCO carrier concentration and accumulation layer thickness indicates the robustness of field effect modulation as a tuning mechanism. The straightforward fabrication of multilayer metamaterials by thin film deposition techniques suggests that that the experimental realization of tunable field effect metamaterials with tunable optical parameters is well within reach.

### Acknowledgments

This work was supported by U.S. Department of Energy (DOE) Office of Science grant DE-FG02-07ER46405 (G.P. and H.A.A.). G. Papadakis acknowledges support by a National Science Foundation Graduate Research Fellowship. We acknowledge fruitful discussions with Prof. Pochi Yeh, Dr. Ho W. Lee, Dr. Krishnan Thyagarajan and Prof. E. N. Economou.

### Author Information

Corresponding author email: gpapadak@caltech.edu

Notes: The authors declare no competing financial interest

# Supplemental Material

## 1. OPTICAL CONSTANT OF Ag (Section II, B)

The "toy model" metamaterial design in the main text incorporates 20nm thick Ag layers. At these length scales, surface effects become important for metallic thin films, thus, we experimentally determine the index of Ag[1]. 20nm thick Ag layers were deposited on Si with e-beam evaporation using an Angstrom Engineering Inc. evaporator. We show in Fig.10 a 5-poles Drude-Lorentz model[2] used for fitting the optical constant of Ag and the results of the fitting. Ellipsometric data were taken with using a J. A. Woollam Co. ellipsometer for angles $40^o$-$80^o$ by increments of $5^o$ and the fitting was performed with the WVASE32 software. The MSE of the fitting was 4.62 and the final thickness was 22.339 nm with a roughness of 0.049±0.05 nm.

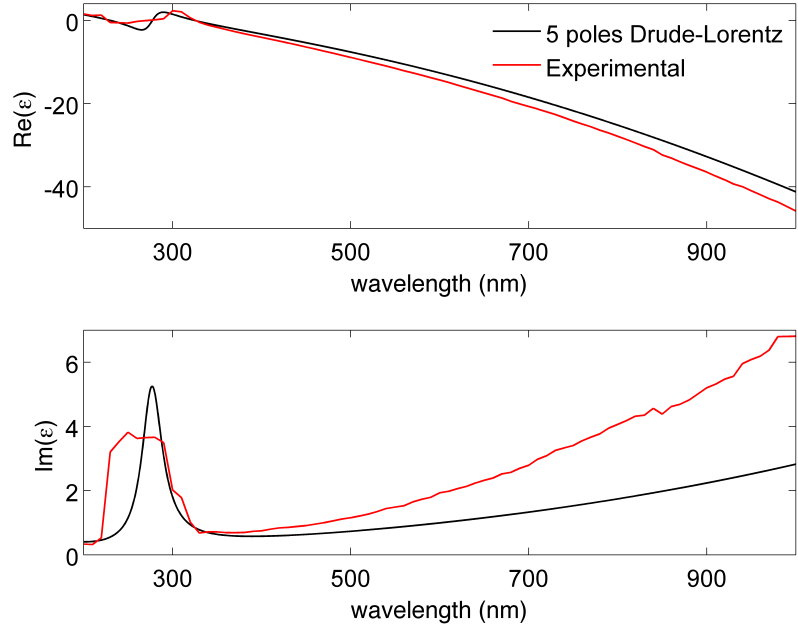

**Fig. 10:** Real (top) and imaginary (bottom) parts of the Ag permittivity. Black color corresponds to a 5-poles Drude Lorentz model used for fitting, red color corresponds to the experimentally determined permittivity

## 2. TUNABLE PERMITTIVITY OF INDIUM TIN OXIDE (Section II, B)

It is well established that in the visible-near IR regime the permittivity of ITO is sufficiently described by the Drude model[3-8].

$$\varepsilon = \varepsilon_\infty - \frac{\omega_p^2}{\omega^2 + i\omega\gamma} \text{ , with } \omega_p^2 = \frac{Ne}{\varepsilon_0 m_{eff}}$$

Based on previous experimental work[4,5,7], the following parameters have been used to model the permittivity of ITO for varying carrier concentration $N$. $\varepsilon_\infty$ which is the high-frequency dielectric permittivity that is taken to be equal to 3.4, $\gamma$ is the electron scattering frequency and is considered here to be equal to $1.8 \cdot 10^{14} / s$. Finally, $m_{eff} = 0.35 m_e$ where $m_e$ is the electronic rest mass.

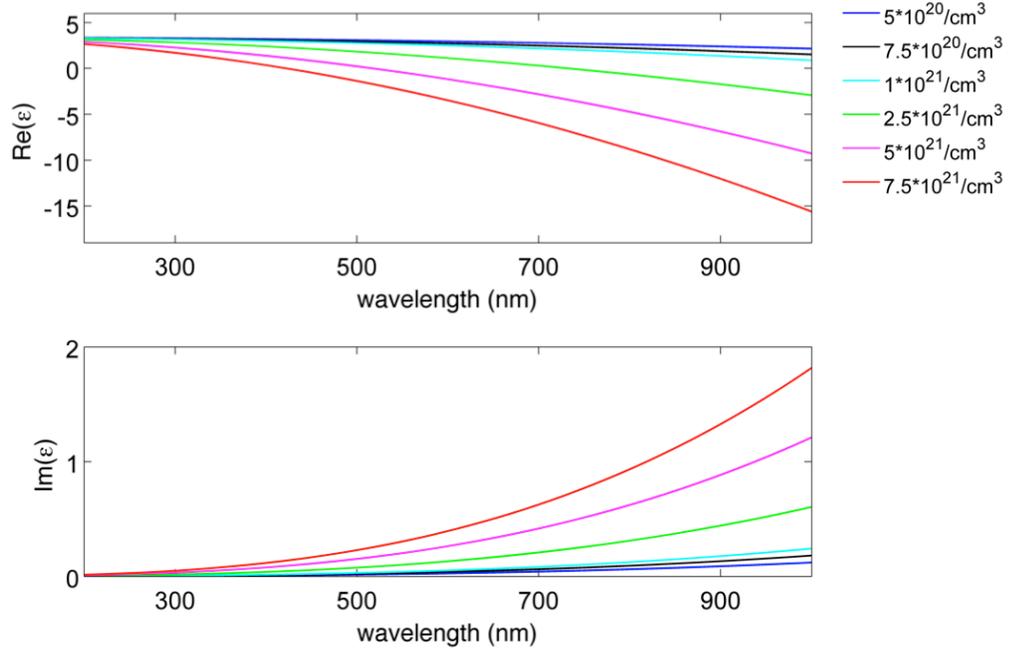

**Fig. 11**: Permittivity of ITO for different considered carrier concentrations

We emphasize here that, relative to previous experimental work of [4] and similar estimates for AZO [9], we underestimate the Debye length of previous experimental findings of 5nm to be 2.5nm for which we perform our calculations. The sensitivity analysis in Part III of the main text shows the trends of the tunable parameters for different Debye lengths.

## 3. LOCALITY OF EFFECTIVE EXTRACTED PARAMETERS

In order to investigate on the locality of the effective parameters, we present in Fig. 12 the extracted extraordinary parameters $\varepsilon_e$ and $\mu_e$ of the considered unit cell metamaterial, for ITO carrier concentration in both accumulation and bulk regions being $5 \times 10^{20} / cm^3$, using the retrieval technique of Ref. 10, for different incident angles. It is noted here that, as proved in Ref. 10, the ordinary parameters $\varepsilon_o$ and $\mu_o$ are only valid for normal incidence and thus, they cannot serve as a metric of locality.

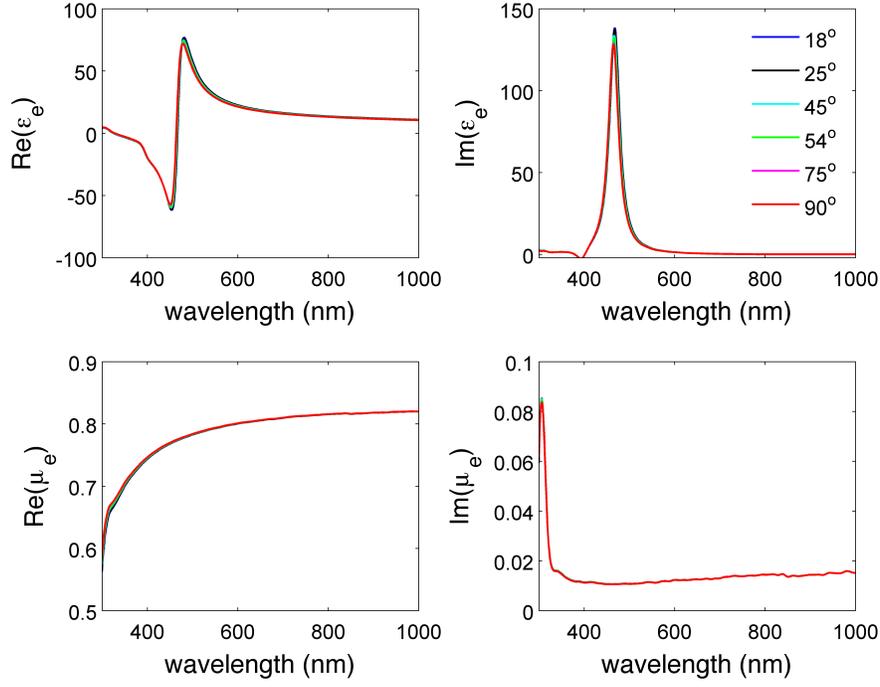

**Fig. 12**: Angle independence of effective extraordinary parameters for carrier concentration $5\times 10^{20}/cm^3$

It can be seen from Fig. 12 that for all the considered incident angles, the extraordinary parameters converge to the same curves. The angle independence of $\varepsilon_e$ and $\mu_e$ implies their independence on the wavevector, and thus, their locality. We obtain similar, angle independent parameters for all the consider carrier concentrations in the ITO accumulation layers. The locality of the effective parameters indicates that any number of unit cell HMMs stacked together yields similar effective parameters to the original unit cell.

## 4. TUNABILITY OF $\varepsilon_o$

We present in Fig. 13 the wavelength dependence of ordinary permittivity $\varepsilon_o$ as the carrier concentration of the accumulation layer ITO increases, for the geometry discussed in the main text.

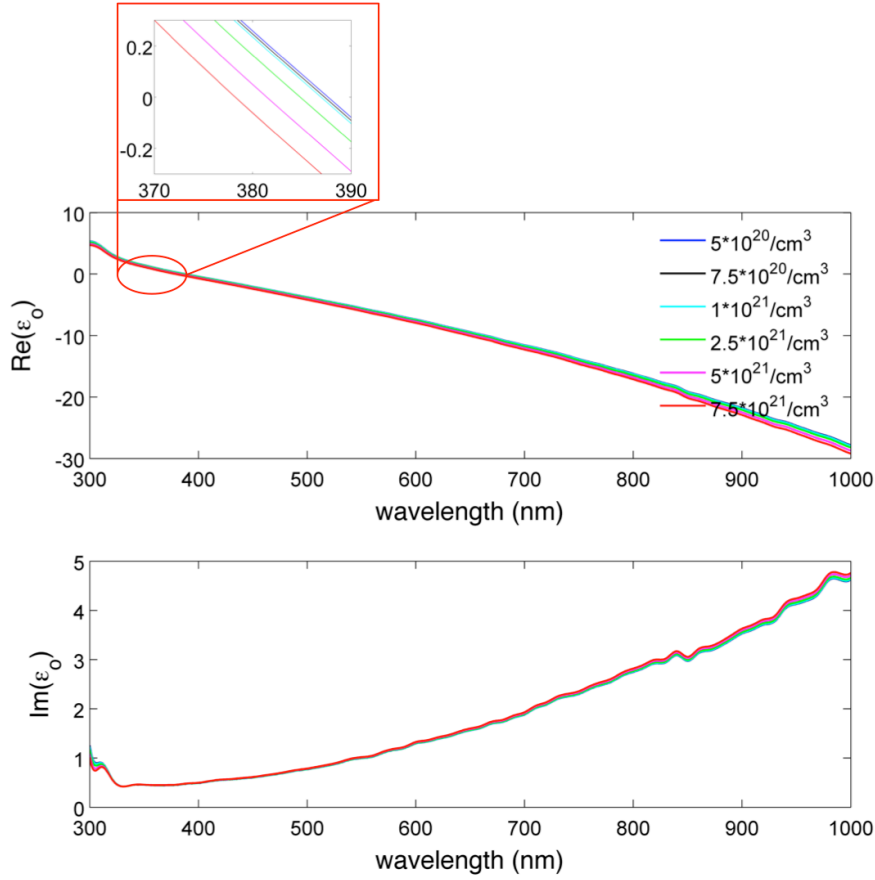

**Fig. 13**: Effective ordinary permittivity. Top: real part and bottom: imaginary part of $\varepsilon_o$ for various carrier concentrations in the accumulation layer

The ordinary permittivity does not exhibit considerable spectral shifting as the carrier concentration increases, which is expected since the Drude-like response in the in-plane direction is mostly affected by the presence of Ag. However, the inset shows tunability in the order of some nanometers of the ENZ wavelength of $\varepsilon_o$ which is sufficient for inducing considerable changes in the dispersion surface of the TE and TM polarized light.

### 4. Tunable figure of merit

Complementary to Fig. 6 of the main text, we present here explicitly the dependence of the figure of merit defined as $FOM = \dfrac{\mathrm{Re}(k_{xeff})}{\mathrm{Im}(k_{xeff})}$ on the carrier concentration in the accumulation layer of ITO. As expected due to the property of HMMs to support large wave numbers, the figures of merit are large compared to other metamaterial realizations. The valleys and ridges of Fig. 6 are clearly reproduced here, which indicates the drastic change in the optical behaviour of the HMM under applied bias.

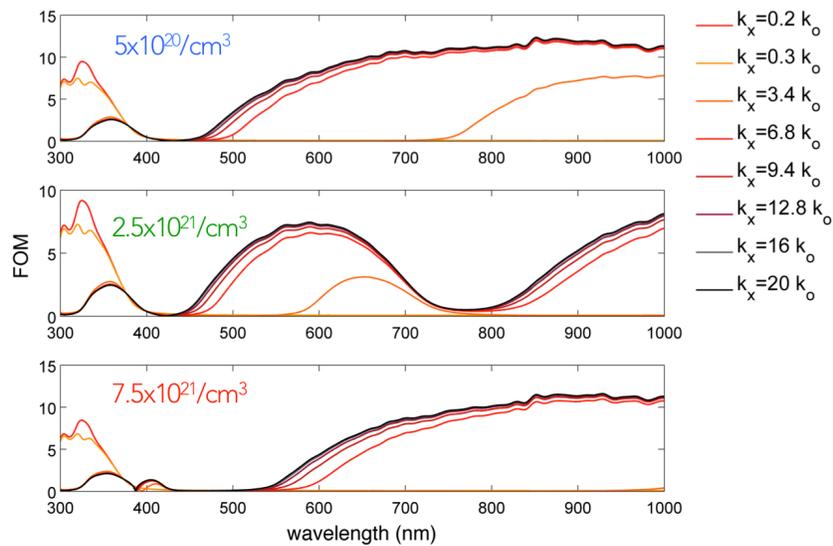

**Fig. 14**: Figure of merit for three different carrier concentrations in the ITO accumulation

**Supplemental Material References**